# The Klein-Gordon-Zakharov equations with the positive fractional power terms and their exact solutions

Jinliang Zhang [†], Wuqiang Hu, Yu Ma

School of Mathematics and Statistics, Henan University of Science and Technology, Luoyang 471023, China

[†]Corresponding author. E-mail: zhangjin6602@163.com

**Abstract:** In this paper, the famous Klein-Gordon-Zakharov equations are firstly generalized, the new special types of Klein-Gordon-Zakharov equations with the positive fractional power terms (gKGZE) are presented. In order to derive the exact solutions of new special gKGZE, the subsidiary higher order ordinary differential equations (sub-ODEs) with the positive fractional power terms are introduced, and with the aids of the Sub-ODE, the exact solutions of three special types of the gKGZE are derived, which are the bell-type solitary wave solution, the algebraic solitary wave solution, the kink-type solitary wave solution and the sinusoidal traveling wave solution, provided that the coefficients of gKGZE satisfy certain constraint conditions.

**Key words:** Klein-Gordon-Zakharov equation with the positive fractional power terms(gKGZE); Sub-ODE method; Exact solution; Constraint condition

## 1. Introduction

In this paper, we consider the famous Klein-Gordon-Zakharov equations:

$$\begin{cases} u_{tt} - k_1^2 u_{xx} + u = nu, \\ n_{tt} - k_2^2 n_{xx} = \left(|u|^2\right)_{xx}. \end{cases} \quad (1)$$

Here in Eqs.(1), the variable $u(x,t)$ is a complex function and $n(x,t)$ is a real function.

Eqs.(1) appear in the area of plasma physics, and they describe the interaction of Langmuir waves and ion-acoustic waves in plasmas[1-3]. Due to their potential application in plasma physics, the nonlinear Klein-Gordon-Zakharov system has been paid attention by many researchers. some exact solutions for the Zakharov equations are obtained using different methods[4-9,14-15]. In the Ref.[4-5], using the F-expansion method, the periodic wave solutions expressed by Jacobi elliptic functions for Klein-Gordon-Zakharov equation are derived. In the Ref.[6], using the extended hyperbolic functions method, the multiple exact explicit solutions of the Klein-Gordon-Zakharov equations are obtained. Using the solitary wave Ansatz method, 1-soliton solution of the Klein-Gordon-Zakharov equation with power law nonlinearity is given, and the numerical simulations are included that supports the analysis[7]. Bifurcation analysis and the travelling wave solutions of the Klein-Gordon-Zakharov equations are studied[8], and topological soliton solution of the Klein-Gordon-Zakharov equation in (1+1)-dimensions with power law nonlinearity are derivrd and bifurcation analysis are studied in Ref.[9]. In the Ref.[14-15], Jacobi elliptic function expansion method is used to derive the periodic solutions for the Klein-Gordon-Zakharov Equations. Gan[10-11] studied the instability of standing waves for Klein-Gordon-Zakharov equations. Linear stability analysis for periodic traveling waves of Klein-Gordon-Zakharov equations are discussed in the Ref.[12-13]. In the Ref.[16-19], the finite difference scheme is proposed for the initial-boundary problem of the Klein-Gordon-Zakharov equations.

The rest of the paper is organized as follows: In Section 2, Klein-Gordon-Zakharov equations are generalized, and the three special types of Klein-Gordon-Zakharov equations with the positive fractional power terms are presented; In Section 3, the sub-ODEs with the positive fractional power terms are introduced, and the exact solutions are presented; In Section 4, the exact solutions of three new special types of the Klein-Gordon-Zakharov equations (5), (6) and (7) are derived with the aids of the sub-ODE with the positive fractional power terms, respectively; In Section 5, some conclusions are made briefly.

## 2. The Klein-Gordon-Zakharov equations with the positive fractional power terms

In the Ref.[20], the Klein-Gordon-Zakharov equations with power law nonlinearity are considered as

$$\begin{cases} u_{tt} - k_1^2 u_{xx} + au + bnu = 0, \\ n_{tt} - k_2^2 n_{xx} = c\left(|u|^{2m}\right)_{xx}. \end{cases} \quad (2)$$

and the soliton solutions are given. It is obvious that the Eqs.(2) becomes Eqs.(1) when we set $a=1, b=-1, c=1, m=1$. And in the Ref.[21], (2+1)-dimensional Klein-Gordon-Zakharov equation with power law nonlinearity are studied as

$$\begin{cases} q_{tt} - \lambda^2\left(q_{xx}+q_{yy}\right) + q + rq + \alpha|q|^{2m}q = 0, \\ r_{tt} - \lambda^2\left(r_{xx}+r_{yy}\right) = \left(|q|^{2m}\right)_{xx} + \left(|q|^{2m}\right)_{yy}, \end{cases} \quad (3)$$

and soliton solutions are presented.

Based on the Refs.[20-21], we generalize the Klein-Gordon-Zakharov equations as

$$\begin{cases} q_{tt} - \lambda^2 \Delta q + aq + brf\left(|q|^2\right)q + \alpha g\left(|q|^2\right)q = 0, \\ r_{tt} - \lambda^2 \Delta r = \beta \Delta\left[h\left(|q|^2\right)\right]. \end{cases} \quad (4)$$

where $f$, $g$ and $h$ are functions of $u$. It is easy to see that the Eqs.(4) becomes Eqs.(3) when we set $a=b=\alpha=\beta=1$, , $\Delta = \partial^2/\partial x^2 + \partial^2/\partial y^2$.

Here when we set $f(u)=g(u)=h(u)=u^{\frac{q}{p}}$ in Eqs.(4), the first new special type of the Klein-Gordon-Zakharov equation with the positive fractional power terms is presented as

$$\begin{cases} u_{tt} - \lambda^2 u_{xx} + au + br|u|^{\frac{q}{p}}u + \alpha|u|^{\frac{q}{p}}u = 0, \\ r_{tt} - \lambda^2 r_{xx} = \beta\left(|u|^{\frac{q}{p}}\right)_{xx}. \end{cases} \quad (5)$$

When we set $f(u)=1$, $g(u)=u^{\frac{q}{p}}+u^{\frac{2q}{p}}$, $h(u)=u^{\frac{q}{p}}$ in Eqs.(4), the second new special type of the Klein-Gordon-Zakharov equation with the positive fractional power terms is presented as

$$\begin{cases} u_{tt} - \lambda^2 u_{xx} + au + bru + \alpha |u|^{\frac{q}{p}} u + \alpha |u|^{\frac{2q}{p}} u = 0, \\ r_{tt} - \lambda^2 r_{xx} = \beta \left( |u|^{\frac{q}{p}} \right)_{xx}. \end{cases} \quad (6)$$

When we set $f(u)=1$, $g(u)=u^{\frac{q}{p}}+u^{\frac{2q}{p}}$, $h(u)=u^{\frac{2q}{p}}$ in Eqs.(4), the third new special type of the Klein-Gordon-Zakharov equation with the positive fractional power terms is presented as

$$\begin{cases} u_{tt} - \lambda^2 u_{xx} + au + bru + \alpha |u|^{\frac{q}{p}} u + \alpha |u|^{\frac{2q}{p}} u = 0, \\ r_{tt} - \lambda^2 r_{xx} = \beta \left( |u|^{\frac{2q}{p}} \right)_{xx}. \end{cases} \quad (7)$$

Our interest in this paper is to derive the exact solutions of these new three special types of the gKGZEs (5)-(7). Because these special types of the gKGZEs (5)-(7) have the positive fractional power nonlinear terms, the sub-ODE with the positive fractional power terms is firstly introduced.

## 3. The sub-ODE with the positive fractional power terms

Inspired by the subsidiary higher order ordinary differential equations[22-29], we consider the nonlinear ODE with the positive fractional power terms as

$$F'^2 = AF^2 + BF^{q/p+2} + CF^{2q/p+2}, \quad (8)$$

where $A$, $B$ and $C$ are constants, $p$ and $q$ are positive integers.

Then Eq.(8) admits exact solutions as follows:

(1) When $A > 0, B = 2\sigma A, C = (\sigma^2 - 1)A, -1 < \sigma < 1$,

$$F = \left[ \frac{1}{\cosh(\sqrt{A}\xi q/p) - \sigma} \right]^{\frac{p}{q}};$$

(2) When $A = 0, B = \dfrac{4p^2}{q^2}, C = -\dfrac{4p^2}{q^2}\sigma, \sigma > 0$,

$$F = \left[\frac{1}{\xi^2 + \sigma}\right]^{\frac{p}{q}};$$

(3) When $A > 0, B = -2\sqrt{AC}, C > 0$,

$$F = \left\{\sqrt{\frac{A}{C}}\left[\frac{1}{2} \pm \frac{1}{2}\tanh\left(\frac{q}{2p}\sqrt{A}\xi\right)\right]\right\}^{\frac{p}{q}};$$

(4) When $A = -1, B = 2\sigma, C = 1 - \sigma^2$,

$$F = \left[\frac{1}{\sigma \pm \sin(q\xi/p)}\right]^{\frac{p}{q}}.$$

Note 1: It should be noted that ODE (8) could admit other solutions, for example, the negative solutions for odd integer $q$, and so on. But for the sake of simplicity, we neglect the cases here.

## 4. Exact solutions of some special types of the Eqs.(4)

### 4.1 The first special type Klein-Gordon-Zakharov equations (5)

Here we suppose the exact solutions of Eqs.(5) are in the form

$$u(x,t) = v(\xi)\exp(i\eta), r(x,t) = r(\xi), \quad \xi = kx - \omega t, \eta = lx - \theta t, \quad (9)$$

where $k, l, \omega, \theta$ are constants determined later.

Substituting (9) into (5) yields nonlinear equations as follows:

$$\theta\omega - lk\lambda^2 = 0, \tag{10}$$

$$(\omega^2 - k^2\lambda^2)v'' + (a + l^2\lambda^2 - \theta^2)v + brv^{\frac{q}{p}+1} + \alpha v^{\frac{q}{p}+1} = 0, \tag{11}$$

$$(\omega^2 - \lambda^2 k^2)r'' - \beta k^2\left(v^{\frac{q}{p}}\right)'' = 0. \tag{12}$$

Integrating (12) twice and setting constants to zero yield

$$r = \frac{\beta k^2}{\omega^2 - \lambda^2 k^2}v^{\frac{q}{p}}. \tag{13}$$

Substituting (13) into (11) yields

$$(\omega^2 - k^2\lambda^2)v'' + (a + l^2\lambda^2 - \theta^2)v + \alpha v^{\frac{q}{p}+1} + \frac{b\beta k^2}{\omega^2 - \lambda^2 k^2}v^{\frac{2q}{p}+1} = 0. \tag{14}$$

Supposing the solutions of (14) are in the form as

$$v = DF(\xi), \tag{15}$$

where $F$ satisfies Eq.(8) and $A, B, C$ and $D$ are constants.

Substituting (15) into (14) and considering Eq.(8) simultaneously, the left-hand side of Eq. (14) becomes a polynomial in $F(\xi)$, when $\omega^2 - k^2\lambda^2 \neq 0$, setting the coefficients of the polynomial in Eq.(14) to zero yields

$$A + \frac{a + l^2\lambda^2 - \theta^2}{\omega^2 - k^2\lambda^2} = 0,$$

$$(\omega^2 - k^2\lambda^2)\left(\frac{q}{2p} + 1\right)B + \alpha D^{\frac{q}{p}} = 0,$$

$$(\omega^2 - k^2\lambda^2)\left(\frac{q}{p} + 1\right)C + \frac{b\beta k^2}{\omega^2 - \lambda^2 k^2} D^{\frac{2q}{p}} = 0,$$

$$\theta\omega - lk\lambda^2 = 0.$$

Solving the algebraic equations obtained above yields

$$A = -\frac{a + l^2\lambda^2 - \theta^2}{\omega^2 - k^2\lambda^2},$$

$$B = -\frac{2p\alpha D^{\frac{q}{p}}}{(\omega^2 - k^2\lambda^2)(q + 2p)},$$

$$C = -\frac{pb\beta k^2}{(\omega^2 - k^2\lambda^2)^2 (q + p)} D^{\frac{2q}{p}}.$$

where $D > 0$, and $\theta\omega - lk\lambda^2 = 0$, $\omega^2 - k^2\lambda^2 \neq 0$.

With the help of the nonlinear sub-ODE with the positive fractional power terms (8), the exact solutions of Eqs.(5) are obtained as

**Case 4.1.1**

$$u(x,t) = \sqrt[q]{\left[\frac{(q+2p)(a+l^2\lambda^2-\theta^2)}{p\alpha}\sigma\right]^p} \left[\frac{1}{\cosh(\sqrt{A}\xi q/p) - \sigma}\right]^{\frac{p}{q}} \exp(i\eta),$$

$$r(x,t) = \frac{\sigma\beta k^2(q+2p)(a+l^2\lambda^2-\theta^2)}{(\omega^2 - \lambda^2 k^2)p\alpha} \frac{1}{\cosh(\sqrt{A}\xi q/p) - \sigma},$$

where $\xi = kx - \omega t, \eta = lx - \theta t$, $\omega = \lambda^2 \dfrac{lk}{\theta}$,

$$\sigma^2 = \dfrac{p\lambda^2\alpha^2(\lambda^2 l^2 - \theta^2)(q+p)}{p\lambda^2\alpha^2(\lambda^2 l^2 - \theta^2)(q+p) - b\beta\theta^2(q+2p)^2(a+l^2\lambda^2 - \theta^2)},$$

and the constants satisfy

$$\dfrac{a+l^2\lambda^2 - \theta^2}{\lambda^2 l^2 - \theta^2} < 0, \dfrac{(a+l^2\lambda^2 - \theta^2)\sigma}{\alpha} > 0, b\beta > 0, \lambda^2 k^2 \left(\dfrac{\lambda^2 l^2}{\theta^2} - 1\right) \neq 0.$$

**Case 4.1.2**

$$u(x,t) = \sqrt[q]{\left[-\dfrac{2p(\omega^2 - k^2\lambda^2)(q+2p)}{\alpha q^2}\right]^p \left[\dfrac{1}{\xi^2 + \sigma}\right]^{\frac{p}{q}}} \exp(i\eta),$$

$$r(x,t) = -\dfrac{2p\beta k^2(\omega^2 - k^2\lambda^2)(q+2p)}{\alpha q^2(\omega^2 - \lambda^2 k^2)} \dfrac{1}{\xi^2 + \sigma},$$

where $\xi = kx - \omega t, \eta = lx - \theta t$, $\omega = \lambda^2 \dfrac{lk}{\theta}, \theta^2 = a + l^2\lambda^2$, $\sigma = \dfrac{pb\beta k^2(q+2p)^2}{\alpha^2(q+p)q^2}$,

and the constants satisfy

$$\dfrac{\lambda^2 l^2 - \theta^2}{\alpha} < 0, b\beta > 0, \text{ and } \lambda^2 k^2\left(\dfrac{\lambda^2 l^2}{\theta^2} - 1\right) \neq 0.$$

**Case 4.1.3**

$$u(x,t) = D\left\{\sqrt{\dfrac{A}{C}}\left[\dfrac{1}{2} \pm \dfrac{1}{2}\tanh\left(\dfrac{q}{2p}\sqrt{A}\xi\right)\right]\right\}^{\frac{p}{q}} \exp(i\eta),$$

$$r(x,t) = \dfrac{\beta k^2}{\omega^2 - \lambda^2 k^2} D^{\frac{q}{p}} \sqrt{\dfrac{A}{C}}\left[\dfrac{1}{2} \pm \dfrac{1}{2}\tanh\left(\dfrac{q}{2p}\sqrt{A}\xi\right)\right],$$

where $\xi = kx - \omega t, \eta = lx - \theta t$,

$$A = -\dfrac{a + l^2\lambda^2 - \theta^2}{\omega^2 - k^2\lambda^2}, C = -\dfrac{pb\beta k^2}{(\omega^2 - k^2\lambda^2)^2(q+p)} D^{\frac{2q}{p}},$$

$$\dfrac{p\alpha^2(\omega^2 - k^2\lambda^2)(q+p)}{b\beta k^2(q+2p)^2} = a + l^2\lambda^2 - \dfrac{l^2 k^2\lambda^4}{\omega^2}, \theta = \dfrac{lk\lambda^2}{\omega},$$

and the constants satisfy

$$\frac{\omega^2\left(a+l^2\lambda^2\right)-l^2k^2\lambda^4}{\omega^2-k^2\lambda^2}<0, b\beta<0, \alpha\left(\omega^2-k^2\lambda^2\right)>0, D>0, \omega^2-k^2\lambda^2\neq 0.$$

**Case 4.1.4**

$$u(x,t)=\sqrt[2q]{\left[\frac{\left(\omega^2-k^2\lambda^2\right)^2(q+2p)^2(q+p)}{p^2\alpha^2(q+p)-pb\beta k^2(q+2p)^2}\right]^p}\left[\frac{1}{\sigma\pm\sin(q\xi/p)}\right]^{\frac{p}{q}}\exp(i\eta),$$

$$r(x,t)=\frac{\beta k^2}{\omega^2-\lambda^2 k^2}\sqrt[q]{\left[\frac{\left(\omega^2-k^2\lambda^2\right)^2(q+2p)^2(q+p)}{p^2\alpha^2(q+p)-pb\beta k^2(q+2p)^2}\right]^p}\frac{1}{\sigma\pm\sin(q\xi/p)},$$

where $\xi=kx-\omega t, \eta=lx-\theta t, \theta^2=k^2\lambda^2\left(1-\dfrac{a}{\omega^2-k^2\lambda^2}\right), l=\dfrac{\theta\omega}{k\lambda^2}$,

$$D=\sqrt[2q]{\left[\frac{\left(\omega^2-k^2\lambda^2\right)^2(q+2p)^2(q+p)}{p^2\alpha^2(q+p)-pb\beta k^2(q+2p)^2}\right]^p},$$

and the constants satisfy

$$\omega^2-k^2\lambda^2\neq 0, \alpha\left(\omega^2-k^2\lambda^2\right)<0, p\alpha^2(q+p)-b\beta k^2(q+2p)^2>0,$$

$$\sigma=-\frac{p\alpha D^{\frac{q}{p}}}{\left(\omega^2-k^2\lambda^2\right)(q+2p)}>1, \frac{a}{\omega^2-k^2\lambda^2}<1.$$

### 4.2 The second special type Klein-Gordon-Zakharov equations (6)

Similar to Section 4.1, supposing the exact solutions of Eqs.(6) are in the form

$$u(x,t)=v(\xi)\exp(i\eta), r(x,t)=r(\xi), \xi=kx-\omega t, \eta=lx-\theta t, \quad (9)$$

where $k,l,\omega,\theta$ are constants determined later. Substituting (9) into (6) yields nonlinear equations as follows:

$$\theta\omega-lk\lambda^2=0,$$

$$r=\frac{\beta k^2}{\omega^2-\lambda^2 k^2}v^{\frac{q}{p}},$$

$$\left(\omega^2-k^2\lambda^2\right)v''+\left(a+l^2\lambda^2-\theta^2\right)v+\left(\frac{\beta k^2 b}{\omega^2-\lambda^2 k^2}+\alpha\right)v^{\frac{q}{p}+1}+\alpha v^{\frac{2q}{p}+1}=0. \quad (16)$$

Supposing the solutions of (15) are in the form as

$$v = DF(\xi), \qquad (15)$$

where $F$ satisfies Eq.(8) and $A, B, C$ and $D$ are constants.

Substituting (15) into (16) and considering Eq. (5) simultaneously, the left-hand side of Eq. (16) becomes a polynomial in $F(\xi)$, when $\omega^2 - k^2\lambda^2 \neq 0$, Setting the coefficients of the polynomial in Eq.(16) to zero yields

$$(\omega^2 - k^2\lambda^2)A + (a + l^2\lambda^2 - \theta^2) = 0,$$

$$\left(\frac{q}{2p} + 1\right)(\omega^2 - k^2\lambda^2)B + \left(\frac{\beta k^2 b}{\omega^2 - \lambda^2 k^2} + \alpha\right)D^{\frac{q}{p}} = 0,$$

$$\left(\frac{q}{p} + 1\right)(\omega^2 - k^2\lambda^2)C + \alpha D^{\frac{2q}{p}} = 0.$$

Solving the algebraic equations obtained above yields

$$A = -\frac{a + l^2\lambda^2 - \theta^2}{\omega^2 - k^2\lambda^2}, \quad B = -\frac{\left(\dfrac{\beta k^2 b}{\omega^2 - \lambda^2 k^2} + \alpha\right)}{\left(\dfrac{q}{2p} + 1\right)(\omega^2 - k^2\lambda^2)} D^{\frac{q}{p}},$$

$$C = -\frac{\alpha D^{\frac{2q}{p}}}{\left(\dfrac{q}{p} + 1\right)(\omega^2 - k^2\lambda^2)},$$

$$\theta\omega - lk\lambda^2 = 0, \omega^2 - k^2\lambda^2 \neq 0, D > 0.$$

With the help of the nonlinear sub-ODE with the positive fractional power terms (8), the exact solutions of Eqs.(6) are obtained as

**Case 4.2.1**

$$u(x,t) = \sqrt[q]{\left\{\frac{(a + l^2\lambda^2 - \theta^2)(q + 2p)(\omega^2 - k^2\lambda^2)\sigma}{p[\beta k^2 b + \alpha(\omega^2 - \lambda^2 k^2)]}\right\}^p \left[\frac{1}{\cosh(\sqrt{A}\xi q/p) - \sigma}\right]^{\frac{p}{q}}} \exp(i\eta),$$

$$r(x,t) = \frac{\beta k^2}{\omega^2 - \lambda^2 k^2} \frac{(a + l^2\lambda^2 - \theta^2)(q + 2p)(\omega^2 - k^2\lambda^2)\sigma}{p[\beta k^2 b + \alpha(\omega^2 - \lambda^2 k^2)]} \frac{1}{\cosh(\sqrt{A}\xi q/p) - \sigma},$$

where $\xi = kx - \omega t, \eta = lx - \theta t$,

$$\sigma^2 = \frac{p(q + p)[\beta k^2 b + \alpha(\omega^2 - \lambda^2 k^2)]^2}{p(q + p)[\beta k^2 b + \alpha(\omega^2 - \lambda^2 k^2)]^2 - \alpha(q + 2p)^2(\omega^2 - k^2\lambda^2)^2},$$

$$\theta = \frac{lk\lambda^2}{\omega},$$

and the constants satisfy

$$\omega^2 - k^2\lambda^2 \neq 0, \ \frac{a + l^2\lambda^2 - \theta^2}{\omega^2 - k^2\lambda^2} < 0, \alpha < 0, \frac{\sigma}{\beta k^2 b + \alpha(\omega^2 - \lambda^2 k^2)} < 0.$$

**Case 4.2.2**

$$u(x,t) = \left[\frac{1}{\xi^2 + \sigma}\right]^{\frac{p}{q}} \sqrt{\left[-\frac{2p(q+2p)(\omega^2 - k^2\lambda^2)^2}{q^2(\beta k^2 b + \alpha(\omega^2 - \lambda^2 k^2))}\right]^p} \exp(i\eta),$$

$$r(x,t) = -\frac{2p\beta k^2(q+2p)(\omega^2 - k^2\lambda^2)^2}{q^2(\omega^2 - \lambda^2 k^2)(\beta k^2 b + \alpha(\omega^2 - \lambda^2 k^2))} \cdot \frac{1}{\xi^2 + \sigma},$$

where $\xi = kx - \omega t, \eta = lx - \theta t$,

$$\sigma = \frac{\alpha p(q+2p)^2(\omega^2 - k^2\lambda^2)^3}{q^2(q+p)(\omega^2 - k^2\lambda^2)(\beta k^2 b + \alpha(\omega^2 - \lambda^2 k^2))^2},$$

$$l = \pm\sqrt{\frac{a\omega^2}{(k^2\lambda^2 - \omega^2)\lambda^2}}, \theta = \pm\frac{k\lambda^2}{\omega}\sqrt{\frac{a\omega^2}{(k^2\lambda^2 - \omega^2)\lambda^2}},$$

and the constants satisfy

$$\omega^2 - k^2\lambda^2 \neq 0, \alpha(\omega^2 - \lambda^2 k^2) > 0, \beta bk^2 + \alpha(\omega^2 - \lambda^2 k^2) < 0, a(k^2\lambda^2 - \omega^2) > 0.$$

**Case 4.2.3**

$$u(x,t) = D\left\{\sqrt{\frac{A}{C}}\left[\frac{1}{2} \pm \frac{1}{2}\tanh\left(\frac{q}{2p}\sqrt{A}\xi\right)\right]\right\}^{\frac{p}{q}} \exp(i\eta),$$

$$r(x,t) = \frac{\beta k^2}{\omega^2 - \lambda^2 k^2} D\sqrt{\frac{A}{C}}\left[\frac{1}{2} \pm \frac{1}{2}\tanh\left(\frac{q}{2p}\sqrt{A}\xi\right)\right],$$

where $\xi = kx - \omega t, \eta = lx - \theta t$,

$$A = -\frac{a + l^2\lambda^2 - \theta^2}{\omega^2 - k^2\lambda^2}, C = -\frac{\alpha D^{\frac{2q}{p}}}{\left(\frac{q}{p} + 1\right)(\omega^2 - k^2\lambda^2)},$$

$$l = \pm\sqrt{\frac{(q+p)p\omega^2}{\alpha(q+2p)^2(\omega^2-k^2\lambda^2)\lambda^2}\left(\frac{\beta k^2 b+\alpha(\omega^2-\lambda^2 k^2)}{\omega^2-\lambda^2 k^2}\right)^2 - \frac{a\omega^2}{(\omega^2-k^2\lambda^2)\lambda^2}},$$

$$\theta = \frac{lk\lambda^2}{\omega},$$

and the constants satisfy

$$\frac{(q+p)p\left(\beta k^2 b+\alpha(\omega^2-\lambda^2 k^2)\right)^2 - a\alpha(q+2p)^2(\omega^2-k^2\lambda^2)^2}{\alpha(\omega^2-k^2\lambda^2)} \geq 0,$$

$$\omega^2 - k^2\lambda^2 \neq 0, \frac{a+l^2\lambda^2-\theta^2}{\omega^2-k^2\lambda^2} < 0, \frac{\alpha}{\omega^2-k^2\lambda^2} < 0, D > 0.$$

**Case 4.2.4**

$$u(x,t) = \left[\frac{1}{\sigma\pm\sin(q\xi/p)}\right]^{\frac{p}{q}} \sqrt[q]{\left[-\frac{\sigma(q+2p)(\omega^2-k^2\lambda^2)^2}{2p^2\left(\beta k^2 b+\alpha(\omega^2-\lambda^2 k^2)\right)}\right]^p} \exp(i\eta),$$

$$r(x,t) = -\frac{\sigma\beta k^2(q+2p)(\omega^2-k^2\lambda^2)^2}{2p^2(\omega^2-\lambda^2 k^2)\left(\beta k^2 b+\alpha(\omega^2-\lambda^2 k^2)\right)} \frac{1}{\sigma\pm\sin(q\xi/p)},$$

where $\xi = kx-\omega t, \eta = lx-\theta t$,

$$\sigma^2 = \frac{4p^3(q+p)\left(\beta k^2 b+\alpha(\omega^2-\lambda^2 k^2)\right)^2}{4p^3(q+p)\left(\beta k^2 b+\alpha(\omega^2-\lambda^2 k^2)\right)^2 - \alpha(q+2p)^2(\omega^2-k^2\lambda^2)},$$

$$l = \pm\sqrt{\frac{\omega^2(\omega^2-k^2\lambda^2-a)}{\lambda^2(\omega^2-k^2\lambda^2)}}, \theta = \pm\frac{k\lambda^2}{\omega}\sqrt{\frac{\omega^2(\omega^2-k^2\lambda^2-a)}{\lambda^2(\omega^2-k^2\lambda^2)}},$$

and the constants satisfy

$$\omega^2 - k^2\lambda^2 \neq 0, \frac{a}{\omega^2-k^2\lambda^2} < 1, \sigma\left[\beta k^2 b+\alpha(\omega^2-\lambda^2 k^2)\right] < 0.$$

### 4.3 The third special type Klein-Gordon-Zakharov equations (4)

Similar to Section 4.1, supposing the exact solutions of Eqs.(7) are in the form

$$u(x,t) = v(\xi)\exp(i\eta), r(x,t) = r(\xi), \xi = kx-\omega t, \eta = lx-\theta t, \tag{15}$$

where $k, l, \omega, \theta$ are constants determined later. Substituting (15) into (7) yields nonlinear equations as follows:

$$\theta\omega - lk\lambda^2 = 0,$$

$$r = \frac{\beta k^2}{\omega^2 - \lambda^2 k^2} v^{\frac{2q}{p}},$$

$$\left(\omega^2 - k^2\lambda^2\right)v'' + \left(a + l^2\lambda^2 - \theta^2\right)v + \alpha v^{\frac{q}{p}+1} + \left(\frac{\beta k^2 b}{\omega^2 - \lambda^2 k^2} + \alpha\right)v^{\frac{2q}{p}+1} = 0. \tag{17}$$

Supposing the solutions of (18) are in the form as

$$v = DF(\xi), \tag{15}$$

where $F$ satisfies Eq.(5) and $A, B, C$ and $D$ are constants.

Substituting (15) into (17) and considering Eq. (8) simultaneously, the left-hand side of Eq. (17) becomes a polynomial in $F(\xi)$, when $\omega^2 - k^2\lambda^2 \neq 0$, Setting the coefficients of the polynomial in Eq.(17) to zero yields

$$\left(\omega^2 - k^2\lambda^2\right)A + \left(a + l^2\lambda^2 - \theta^2\right) = 0,$$

$$\left(\frac{q}{2p} + 1\right)\left(\omega^2 - k^2\lambda^2\right)B + \alpha D^{\frac{q}{p}} = 0,$$

$$\left(\frac{q}{p} + 1\right)\left(\omega^2 - k^2\lambda^2\right)C + \left(\frac{\beta k^2 b}{\omega^2 - \lambda^2 k^2} + \alpha\right)D^{\frac{2q}{p}} = 0.$$

Solving the algebraic equations obtained above yields

$$A = -\frac{a + l^2\lambda^2 - \theta^2}{\omega^2 - k^2\lambda^2},$$

$$B = -\frac{2p\alpha D^{\frac{q}{p}}}{(q + 2p)\left(\omega^2 - k^2\lambda^2\right)},$$

$$C = -\frac{p\left(\beta k^2 b + \alpha\left(\omega^2 - \lambda^2 k^2\right)\right)D^{\frac{2q}{p}}}{(q + p)\left(\omega^2 - k^2\lambda^2\right)^2},$$

$$D > 0, \omega^2 - k^2\lambda^2 \neq 0.$$

With the help of the nonlinear sub-ODE with the positive fractional power terms (8), the exact solutions of Eqs.(17) are obtained as

**Case 4.3.1**

$$u(x,t) = \sqrt[q]{\left[\frac{(q+2p)(a+l^2\lambda^2-\theta^2)}{p\alpha}\sigma\right]^p \left[\frac{1}{\cosh(\sqrt{A}\xi q/p)-\sigma}\right]^{\frac{p}{q}}} \exp(i\eta),$$

$$r(x,t) = \frac{\beta k^2}{\omega^2-\lambda^2 k^2}\left(\frac{\sigma(q+2p)(a+l^2\lambda^2-\theta^2)}{p\alpha(\cosh(\sqrt{A}\xi q/p)-\sigma)}\right)^2,$$

where $\xi = kx - \omega t, \eta = lx - \theta t$,

$$\sigma^2 = \frac{1}{1 - \dfrac{(q+2p)^2(a+l^2\lambda^2-\theta^2)(\beta k^2 b + \alpha(\omega^2-\lambda^2 k^2))}{p\alpha^2(q+p)(\omega^2-k^2\lambda^2)}},$$

$$A = -\frac{a+l^2\lambda^2-\theta^2}{\omega^2-k^2\lambda^2},$$

and the constants satisfy

$$\omega^2 - k^2\lambda^2 \neq 0, \frac{a+l^2\lambda^2-\theta^2}{\omega^2-k^2\lambda^2} < 0, \frac{(a+l^2\lambda^2-\theta^2)\sigma}{\alpha} > 0, \beta k^2 b + \alpha(\omega^2-\lambda^2 k^2) > 0.$$

**Case 4.3.2**

$$u(x,t) = \sqrt[q]{\left[-\frac{2p(q+2p)(\omega^2-k^2\lambda^2)}{\alpha q^2}\right]^p \left[\frac{1}{\xi^2+\sigma}\right]^{\frac{p}{q}}} \exp(i\eta),$$

$$r(x,t) = \frac{\beta k^2}{\omega^2-\lambda^2 k^2}\left(-\frac{2p(q+2p)(\omega^2-k^2\lambda^2)}{\alpha q^2}\cdot\frac{1}{\xi^2+\sigma}\right)^2,$$

where $\xi = kx - \omega t, \eta = lx - \theta t,$,

$$\sigma = \frac{p(q+2p)^2(\beta k^2 b + \alpha(\omega^2-\lambda^2 k^2))}{\alpha^2 q^2(q+p)},$$

and the constants satisfy

$$a + l^2\lambda^2 - \theta^2 = 0, \beta k^2 b + \alpha(\omega^2-\lambda^2 k^2) > 0, \omega^2 - k^2\lambda^2 \neq 0$$

$$\frac{\omega^2 - k^2\lambda^2}{\alpha} < 0.$$

**Case 4.3.3**

$$u(x,t) = D\left\{\sqrt{\frac{A}{C}}\left[\frac{1}{2} \pm \frac{1}{2}\tanh\left(\frac{q}{2p}\sqrt{A}\xi\right)\right]\right\}^{\frac{p}{q}} \exp(i\eta),$$

$$r(x,t) = \frac{\beta k^2 D^{\frac{2q}{p}}}{\omega^2 - \lambda^2 k^2}\left\{\sqrt{\frac{A}{C}}\left[\frac{1}{2} \pm \frac{1}{2}\tanh\left(\frac{q}{2p}\sqrt{A}\xi\right)\right]\right\}^{2},$$

where $\xi = kx - \omega t, \eta = lx - \theta t, \omega^2 - k^2\lambda^2 \neq 0$,

$$A = -\frac{a + l^2\lambda^2 - \theta^2}{\omega^2 - k^2\lambda^2}, C = -\frac{p\left(\beta k^2 b + \alpha\left(\omega^2 - \lambda^2 k^2\right)\right)D^{\frac{2q}{p}}}{(q+p)\left(\omega^2 - k^2\lambda^2\right)^2},$$

$$\theta^2 = a + l^2\lambda^2 - \frac{p\left(\omega^2 - k^2\lambda^2\right)(q+p)\alpha^2}{(q+2p)^2\left(\beta k^2 b + \alpha\left(\omega^2 - \lambda^2 k^2\right)\right)},$$

and the constants satisfy

$$D > 0, \frac{a + l^2\lambda^2 - \theta^2}{\omega^2 - k^2\lambda^2} < 0, \frac{\alpha}{\left(\omega^2 - k^2\lambda^2\right)} > 0,$$

$$\beta k^2 b + \alpha\left(\omega^2 - \lambda^2 k^2\right) < 0.$$

**Case 4.3.4**

$$u(x,t) = \left[-\frac{\sigma(q+2p)\left(\omega^2 - k^2\lambda^2\right)}{p\alpha\left(\sigma \pm \sin(q\xi/p)\right)}\right]^{\frac{p}{q}} \exp(i\eta),$$

$$r(x,t) = \frac{\beta k^2}{\omega^2 - \lambda^2 k^2}\left(-\frac{\sigma(q+2p)\left(\omega^2 - k^2\lambda^2\right)}{p\alpha\left(\sigma \pm \sin(q\xi/p)\right)}\right)^{2},$$

where $\xi = kx - \omega t, \eta = lx - \theta t$,

$$\omega^2 = a + l^2\lambda^2 - \theta^2 + k^2\lambda^2, \sigma^2 = \frac{1}{1 - \frac{(q+2p)^2\left(\beta k^2 b + \alpha a + \alpha l^2\lambda^2 - \alpha\theta^2\right)}{p\alpha^2(q+p)}},$$

and the constants satisfy

$$a + l^2\lambda^2 - \theta^2 > 0, \frac{\sigma}{\alpha} < 0,$$

$$\frac{(q+2p)^2\left(\beta k^2 b + a\alpha + l^2\lambda^2\alpha - \alpha\theta^2\right)}{p\alpha^2(q+p)} < 1, a \neq \theta^2 - l^2\lambda^2.$$

Note 2: If the $f(u)$, $g(u)$ and $h(u)$ are setted to other special functions, the exact solutions of Klein-Gordon-Zakharov equations with the positive fractional power terms can be obtained. Here we omitted for simplicity.

Note 3: The exact solutions of multidimensional Klein-Gordon-Zakharov equations with the positive fractional power terms can be obtained by the methods using in this paper. Here we omitted for simplicity.

## 5. Conclusions and Discussions

In this paper, the Klein-Gordon-Zakharov equations are firstly generalized, the special types of Klein-Gordon-Zakharov equations with the positive fractional power terms are introduced.

Secondly, the subsidiary higher order ordinary differential equations with the positive fractional power terms are presented, and can be used to derive the exact solutions of the nonlinear evolution equations with the positive fractional power terms.

Thirdly, the exact solutions of three special types of the Klein-Gordon-Zakharov equations with the positive fractional power terms are derived with the aids of the Sub-ODE, which are the bell-type solitary wave solution, the algebraic solitary wave solution, the kink-type solitary wave solution and the sinusoidal traveling wave solution, provided that the coefficients of gKGZE satisfy certain constraint conditions.

It is obvious that the method introduced in this paper may be applied to explore the exact solutions for the other nonlinear evolution equations with the positive fractional power terms.

## Acknowledgement

The authors would like to express the sincere thanks to the referees for their valuable suggestions. This project is supported in part by the Basic Science and the Front Technology Research Foundation of Henan Province of China (Grant No. 092300410179); and the Scientific Research Innovation Ability Cultivation Foundation of Henan University of Science and Technology (Grant no.011CX011).